\documentclass[twocolumn,showpacs,prl]{revtex4}
\usepackage{amssymb}
\usepackage{amsmath}
\usepackage{graphicx}
\usepackage{bm}

\newcommand{\w}{\ensuremath{\omega}}

\begin{document}

\title{Optical emulation of double-continuum Fano interference by circularly dichroic plasmonic metasurfaces}
\author{Nihal Arju, Tzuhsuan Ma, Alexander Khanikaev, David Purtseladze, and Gennady Shvets}
\affiliation{Department of Physics
and Institute for Fusion Studies, The University of Texas at
Austin, One University Station C1500, Austin, Texas 78712}
\date{\today}

\begin{abstract}
Classical emulation of a ubiquitous quantum mechanical phenomenon
of double-continuum Fano (DCF) interference using metasurfaces is
experimentally realized by engineering the near-field interaction
between two bright and one dark plasmonic modes. The competition
between the bright modes, with one of them effectively suppressing the Fano interference for the orthogonal light polarization, is
discovered and explained by the analytic theory of the plasmonic
DCF interference, which is further applied to predicting the
circularly dichroic optical field concentration by plasmonic
metasurfaces.
\end{abstract}


\maketitle

Ugo Fano's breakthrough paper~\cite{fano} that explained
asymmetric ionization spectra in atomic systems by introducing the
now eponymous interference between discrete and continuous states
continues to be highly influential despite its publication more
than half a century ago. Universal concepts of Fano resonance,
interference, and lineshape have been applied to several areas of
optical science, including photonics, plasmonic nanostructures,
and metamaterials
~\cite{fan_prb02,fedotov_srr_fano,mirosh_rmp02,giessen_natmat,lukyan_nmat10}.
The sharp spectral features in Fano-resonant metamaterials,
combined with strong optical field concentration, make them
attractive for sensing/fingerprinting
~\cite{norlander_fano_sensing,giessen_single_molc,guo_fano_sensing,wu_fano_nmat}
and nonlinear~\cite{pendry_field_enhance,benninnk_field_enhance}
applications.

Most of the recently
reviewed~\cite{mirosh_rmp02,lukyan_nmat10,khan_nanophot13} work on
optical Fano resonances deals with near-field coupling between a
single bright and a single dark resonances, respectively emulating
the continuum and discrete atomic states. However, the original
Fano paper addressed a much broader class of couplings between
atomic states, including one discrete and multiple continuum
states. The key difference between the single-continuum and
double-continuum cases is that the ionization probability vanishes
for at least some values of energy loss of the ionizing particle
in the former but not in the latter case~\cite{fano}. Thus, the
magnitude of Fano interference can be suppressed by the second
continuum state.

\begin{figure}[ht]
\includegraphics[width=0.85\columnwidth]{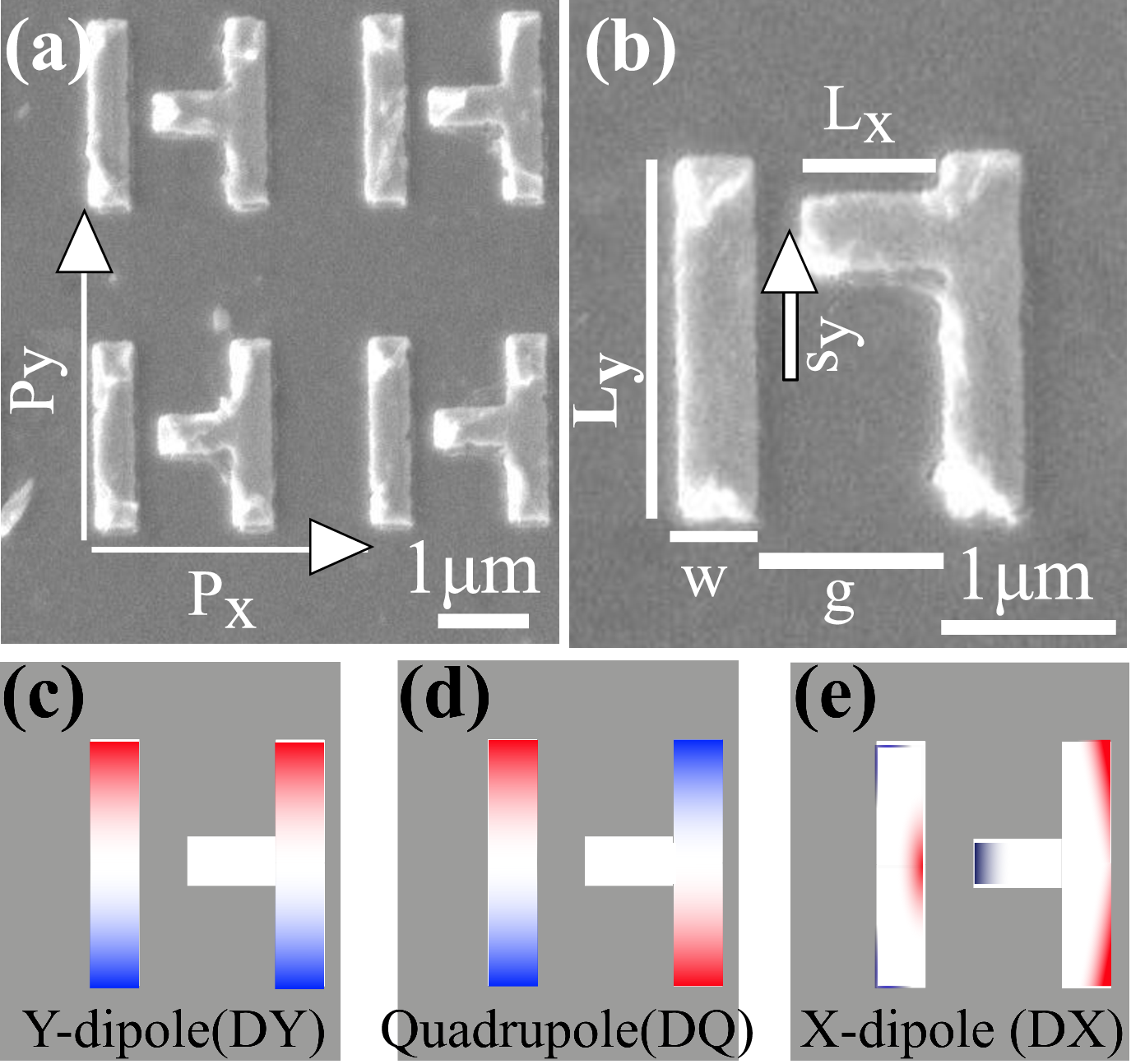}
\vspace{0.11in}
\caption{(color online) (a) Scanning electron micrograph
of a fabricated DCF metasurface. (b) Geometry definitions
of a unit cell ($s_y=0.52\mu$m). For all metasurfaces:
$P_x=2.7\mu m$, $P_y=3.15\mu m$, $w=0.36\mu m$, $L_x=0.92\mu m$,
$L_y=1.8\mu m$, and $g=0.66\mu m$. (d-f): Charge distributions
of the three unperturbed eigenmodes of the symmetric ($s_y=0$)
metasurface. (d) and (f): bright DY and DX, (e): dark DQ.}\label{fig:basis}
\end{figure}

\begin{figure}[ht]
\includegraphics[width=0.95\columnwidth]{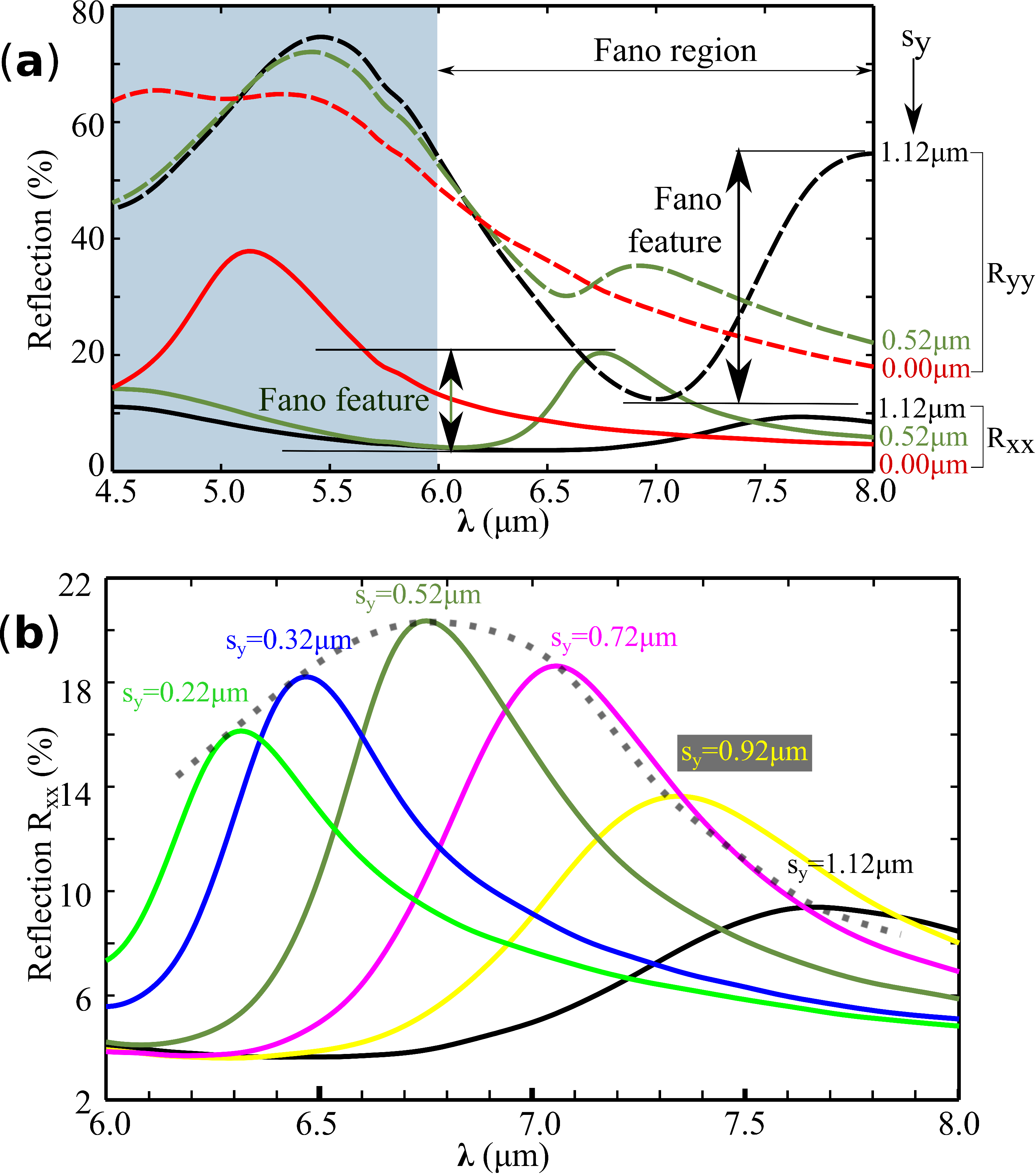} 
\caption{(color online) (a) Measured reflectivity spectra $R_{xx}$ (solid lines) and $R_{yy}$ (dashed lines). Fano feature in $R_{yy}$ is larger than in $R_{xx}$ for $s_y=1.12 \mu$m (black line); opposite is true for $s_y=0.52 \mu$m (green line). Fano features are absent for $s_y=0$ (red lines). (b) Zoomed-in $R_{xx}$ spectra for $0.22\mu$m$< s_y <1.22\mu$m. Dotted line: non-monotonic behavior of the Fano feature with $s_y$ interpreted as the competition between $x-$ and $y-$polarized continua.}\label{fig:spectra}
\end{figure}

In this Letter, we report the realization of the optical analog of
double-continuum Fano (DCF) interference using a circularly
dichroic (CD) plasmonic metasurface shown in Fig.~\ref{fig:basis}
which supports one dark and two bright plasmonic modes. The modes
are controllably coupled to each in the near field through
symmetry-breaking vertical displacement $s_y$ of a horizontal
nanorod coupler (HNC) from its symmetric position. Our
experimental measurements of Fano-shaped polarized reflectivity
spectra $R_{xx}(\lambda)$ and $R_{yy}(\lambda)$ shown in
Fig.~\ref{fig:spectra}(a,b) reveal a new optical phenomenon of
{\it continuum state competition} in asymmetric photonic
structures: the presence of the second (e.g., $y-$polarized)
continuum state can significantly affect the strength of the Fano
interference of the dark state with the first (e.g.,
$x-$polarized) continuum state. Unlike the plasmonic
phenomenon~\cite{martin_prb11} of the Fano feature reduction by
{\it non-radiative} (Ohmic) losses, the continuum state
competition directly emulates the atomic systems~\cite{fano} where the non-radiative decay rate of the discrete state is naturally
negligible. This phenomenon was not addressed in the earlier optical DCF work~\cite{wu_prl11,panoiu_prl13} because the relative mode coupling cannot be controlled in a fixed geometry structure.

{\it Experimental results}-- A family of gold metasurfaces shown
in Fig.~\ref{fig:basis} with different symmetry-breaking
parameters $s_y$ were fabricated on CaF$_2$ substrates using
electron beam lithography. The lowest plasmonic resonances of the
metasurfaces shown in Fig.~\ref{fig:basis} are classified as one
dark quadrupolar (DQ) and two bright dipolar (DX and DY) plasmonic
modes that are radiatively coupled to incident light polarized in
the $x-$  and $y-$directions, respectively, giving rise to two
quasi-continua of electromagnetic states. The finite displacement
$s_y$ of the HNC that perturbs the DQ mode is used to control the
coupling between the modes. Polarized reflection coefficients were
measured using Fourier transform infrared micro-spectroscopy. The
$R_{yy}(\lambda)$ and $R_{xx}(\lambda)$ spectra shown in
Fig.~\ref{fig:spectra}(a) reveal broad reflectivity peaks inside
the shaded ($4.5\mu m < \lambda < 6\mu m$) spectral window,
corresponding to the bright DY and DX resonances, respectively. In
the rest of the Letter we concentrate on the $6\mu m <
\lambda_{DQ}(s_y) < 8\mu m$ spectral window containing the dark DQ
mode.

Several key observations can be made based on the spectral Fano features measured for a wide range of the symmetry-breaking parameter $s_y$. {\it First,} we conclude from Fig.~\ref{fig:spectra}(a) that the relative coupling strengths $\tilde{\kappa}_{XQ}(s_y)$ (between DQ and DX)
and $\tilde{\kappa}_{YQ}(s_y)$ (between DQ and DY) vary widely with $s_y$. This conjecture follows from the very different manifestations of Fano resonance in $R_{xx}$ and $R_{yy}$ spectra. Specifically, for small values of $s_y < L_y/4$, the Fano feature in the $x-$polarization becomes strongly pronounced while it is barely noticeable in $y-$polarization (e.g., $s_y=0.52\mu$m). The opposite is true for large values of $s_y$: the Fano feature in $R_{yy}(\lambda)$ is much stronger than in $R_{xx}(\lambda)$ for $s_y \geq L_y/2$ (e.g., $s_y=1.12\mu$m).

{\it Second,} we observe from Fig.~\ref{fig:spectra}(b) that the magnitude of the Fano feature in $x$-polarization is non-monotonic in $s_y$. The peak value of $R_{xx}$ (dotted line) plotted in Fig.~\ref{fig:spectra}(b) increases with $s_y$, reaches the maximum at $s_y \equiv s_y^{\rm max} \approx 0.52\mu$m, and decreases for $s_y > s_y^{\rm max}$. The decrease of the Fano feature for large values of $s_y$ is unexpected because the coupling coefficient $\tilde{\kappa}_{XQ}(s_y)$ continues monotonically increasing with $s_y$ even for $s_y > s_y^{\rm max}$. The observed decrease is the direct experimental evidence of continuum state competition in DCF originally predicted by
Fano~\cite{fano} for atomic systems. Specifically, the linewidth $\delta \lambda_{DQ}(s_y)$ of the radiatively-broadened Fano resonance increases (and the corresponding quality factor $Q(s_y) = \lambda_{DQ}/(\delta \lambda_{DQ})$ decreases) faster with $s_y$ than the coupling coefficient $\tilde{\kappa}_{XQ}(s_y)$, thereby
suppressing the Fano feature in $R_{xx}$.

{\it Theoretical model of DCF in plasmonic metasurfaces}-- To explain these experimental results and to explore the possibility of circularly dichroic optical field concentration, in what follows we develop a simple analytic model of optical DCF. The interaction of light with the three modes (two quasi-continuum and one discrete)
of the metasurface is described by the following equations based
on the~\cite{haus,joannop_josa03} coupled mode theory (CMT):
\begin{eqnarray}
  \frac{d D_Y}{dt} &=& i \tilde{\omega}_Y D_Y + i \kappa_{XY} D_X +
  \alpha_y E_y^{in} \nonumber \\
  \frac{d Q}{dt} &=& i \tilde{\omega}_Q Q + i \kappa_{XQ} D_X \nonumber \\
  \frac{d D_X}{dt} &=& i \tilde{\omega}_X D_X +  i \kappa_{XY} D_Y +
  i \kappa_{XQ} Q + \alpha_x E_x^{in},
  \label{eqn:cmt2}
\end{eqnarray}
where $D_X$, $D_Y$, and $Q$ are the mode amplitudes, and
$\alpha_{x(y)}$ are the radiative coupling efficiencies of the
bright resonances to the far-field $x(y)-$polarized incident waves
with amplitudes $E_{x(y)}^{in}$, respectively. The three modes are
characterized by their complex-valued unperturbed eigenfrequencies
$\tilde{\omega}_{X,Q,Y} \equiv \omega_{m} - i \tau^{-1}_{m}$
($m=X,Y,Q$ is the resonance's index), where $\omega_{m}$ and
$\tau_{m}$ are the spectral position and unperturbed lifetime of
the $m'$th mode, respectively.

Note that the $\kappa_{YQ}$ component of the near-field coupling
tensor $\kappa_{lm}$ has been neglected in Eq.(\ref{eqn:cmt2}) in
comparison with its $\kappa_{XQ}(s_y)$ and $\kappa_{XY}(s_y)$
components for small value of $s_y$. Qualitatively, $\kappa_{lm}$
is proportional to the overlap integral $\vec{E}_l \cdot
\vec{E}_m$ inside the region occupied by the HNC. In can be
observed from Fig.~\ref{fig:basis}(d-f) that (a) $E_x$ is the
largest electric field component for all three modes in the said
region of space, and (b) $E_{x}=0$ at $y=0$ for the DQ and DY
modes. Therefore, it is expected that $\kappa_{XQ},\kappa_{XQ}
\propto s_y$, but $\kappa_{YQ} \propto s_y^2$ can be neglected for
$s_y \ll L_y/4$. Also, from the energy
conservation~\cite{haus,FanJPCFano,khan_nanophot13}, $1/\tau_{X,Y}
= |\alpha_{x,y}|^2 + 1/\tau_{X,Y}^{\rm Ohm}$, where
$1/\tau_{X,Y}^{\rm Ohm}$ are the intrinsic decay rates.

After solving Eq.(\ref{eqn:cmt2}) in the vicinity of the dark
resonance under the small-coupling ($\left| \kappa_{XY} \right|,
\left| \kappa_{XQ} \right| \ll \left| \omega_Q - \omega_{X,Y}
\right| $) assumption and using the expressions for the
complex-valued reflection amplitudes~\cite{khan_nanophot13}
$r_{xx(yy)} = \alpha^{*}_{x(y)} D_{X(Y)}/E_{x(y)}^{in}$ , we
obtain:
\begin{eqnarray}\label{eqn:r}
  r_{xx} &=& \frac{\alpha_x^2}{(\omega - \tilde{\omega}_X)} +
  \frac{\alpha_x^2 \tilde{\kappa}_{XY}^2}{(\omega - \tilde{\omega}_Y)} + \frac{\alpha_x^2 \tilde{\kappa}_{XQ}^2}{(\omega - \tilde{\omega}_Q^{\prime})}, \label{eq:rxx} \\
  r_{yy} &=& \frac{\alpha_y^2}{(\omega - \tilde{\omega}_Y)} +
  \frac{\alpha_y^2 \tilde{\kappa}_{XY}^2}{(\omega - \tilde{\omega}_X)} + \frac{\alpha_y^2 \tilde{\kappa}_{YQ}^2}{(\omega - \tilde{\omega}_Q^{\prime})} \label{eq:ryy},
\end{eqnarray}
where the normalized coupling coefficients $\tilde{\kappa}_{lm}$
between the modes are given by
\begin{equation}\label{eq:kappa_YQ}
    \tilde{\kappa}_{XY} = \frac{\kappa_{XY}}{\tilde{\omega}_X - \tilde{\omega}_Y},
    \tilde{\kappa}_{XQ} = \frac{\kappa_{XQ}}{\tilde{\omega}_X - \tilde{\omega}_Q},
    \tilde{\kappa}_{YQ} = \tilde{\kappa}_{XY} \tilde{\kappa}_{XQ}.
\end{equation}
The renormalized/red-shifted frequency $\omega_Q^{\prime}$ and the radiatively reduced lifetime $\tau^{\prime}_Q$ of the DQ mode are approximated as
\begin{eqnarray}
  &&\omega^{\prime}_Q \approx  \omega_Q - \frac{\kappa_{XQ}^2}{(\omega_X - \omega_Q)} -
  \frac{\kappa_{XY}^2 \kappa_{XQ}^2}{(\omega_X - \omega_Q)^2 (\omega_Y - \omega_Q)},\nonumber \\
  &&\frac{1}{\tau^{\prime}_Q} \approx \frac{1}{\tau_Q^{\rm Ohm}} + \kappa_{XQ}^2
  \frac{\alpha _x^2}{(\w_Q-\w_X)^2}  + \kappa_{XY}^2 \kappa_{XQ}^2 \times \nonumber \\
  &&\frac{\alpha_y^2 (\w_X-\w_Q)^2 + 2\alpha _x^2 (\w_X-\w_Q)
  (\w_Y-\w_Q)}{(\w_X-\w_Q)^4 (\w_Y-\w_Q)^2}, \label{eq:q_factor}
\end{eqnarray}
where the large modal separation assumption of
$|\omega_{X,Y} - \omega_Q| \gg 1/\tau_{X,Y}$ is used.

The two key features of the experimentally measured $R_{xx(yy)}
\equiv |r_{xx(yy)}|^2$ reflectivity spectra can now be understood
by examining the dependence of the Fano feature's magnitude
$r_{xx(yy)}^{\rm Fano}$ on $s_y$. It is given by the third term in
the rhs of Eqs.~(\ref{eq:rxx},\ref{eq:ryy}) evaluated at
$\omega=\omega^{\prime}_Q$:
\begin{equation}\label{eq:rxx_Fano}
    r_{xx}^{\rm Fano} \propto \frac{\alpha_x^2 s_y^2}{1/\tau^{\rm Ohm}_Q +
    \beta s_y^2 + \gamma s_y^4}, r_{yy}^{\rm Fano} \propto \frac{\alpha_y^2 s_y^4}{1/\tau^{\rm Ohm}_Q +
    \beta s_y^2 + \gamma s_y^4},
\end{equation}
where $\beta$ and $\gamma$ follow from Eq.~(\ref{eq:q_factor}).

We observe that the decay rate of the DCF resonance given by the
denominators of Eq.(\ref{eq:rxx_Fano}) can be broken up into three contributions: (a) the Ohmic (non-radiative) contribution, (b) the contribution $\propto s_y^2$ corresponding to the radiative decay into the x-polarized continuum, and (c) the contribution $\propto s_y^4$ corresponding to the radiative decay into the y-polarized continuum. Depending on the relative dominance of these three mechanisms controlled by $s_y$, the three respective coupling regimes can be identified as (i) weak coupling regime ($s_y <L_y/4$), (ii) intermediate coupling regime ($L_y/4 < s_y \ll L_y/2$), and (iii) strong coupling regime ($s_y \sim L_y/2$). The transition from weak to strong coupling regimes with increasing $s_y$ explains the decrease of the quality factor of the Fano resonance with $s_y$ evident from Fig.~\ref{fig:spectra}(b). The experimentally estimated quality-factors of the DQ mode drop from its Ohmic loss limited value~\cite{giessen_natmat} of $Q=13$ to $Q=7$ (strong $y-$polarized radiative loss) as $s_y$ increases from $s_y=0.22\,\mu m$ to $s_y=1.12 \,\mu m$.

The slower emergence of the Fano feature in the $R_{yy}$ spectrum
compared with the $R_{xx}$ ({\it first} observation)
can be understood from the scaling of $r_{yy}^{\rm Fano} \propto
s_y^4$ versus $r_{xx}^{\rm Fano} \propto s_y^2$ in the weak
coupling regime. The situation changes dramatically for $s_y \geq
L_y/2$ and the Fano feature in $R_{yy}$ becomes stronger than in $R_{xx}$ because of the $\alpha_y > \alpha_x$ relationship which is the consequence of $L_y > L_x$. The {\it second} observation is also explained by the non-monotonic dependence of $r_{xx}^{\rm Fano}$ on $s_y$ due to the transition to strong coupling regime, where the continuum state competition~\cite{fano} in DCF is responsible for the weakening of the Fano feature in $R_{xx}$ for large $s_y$.

\begin{figure}[ht]
\includegraphics[width=0.85\columnwidth]{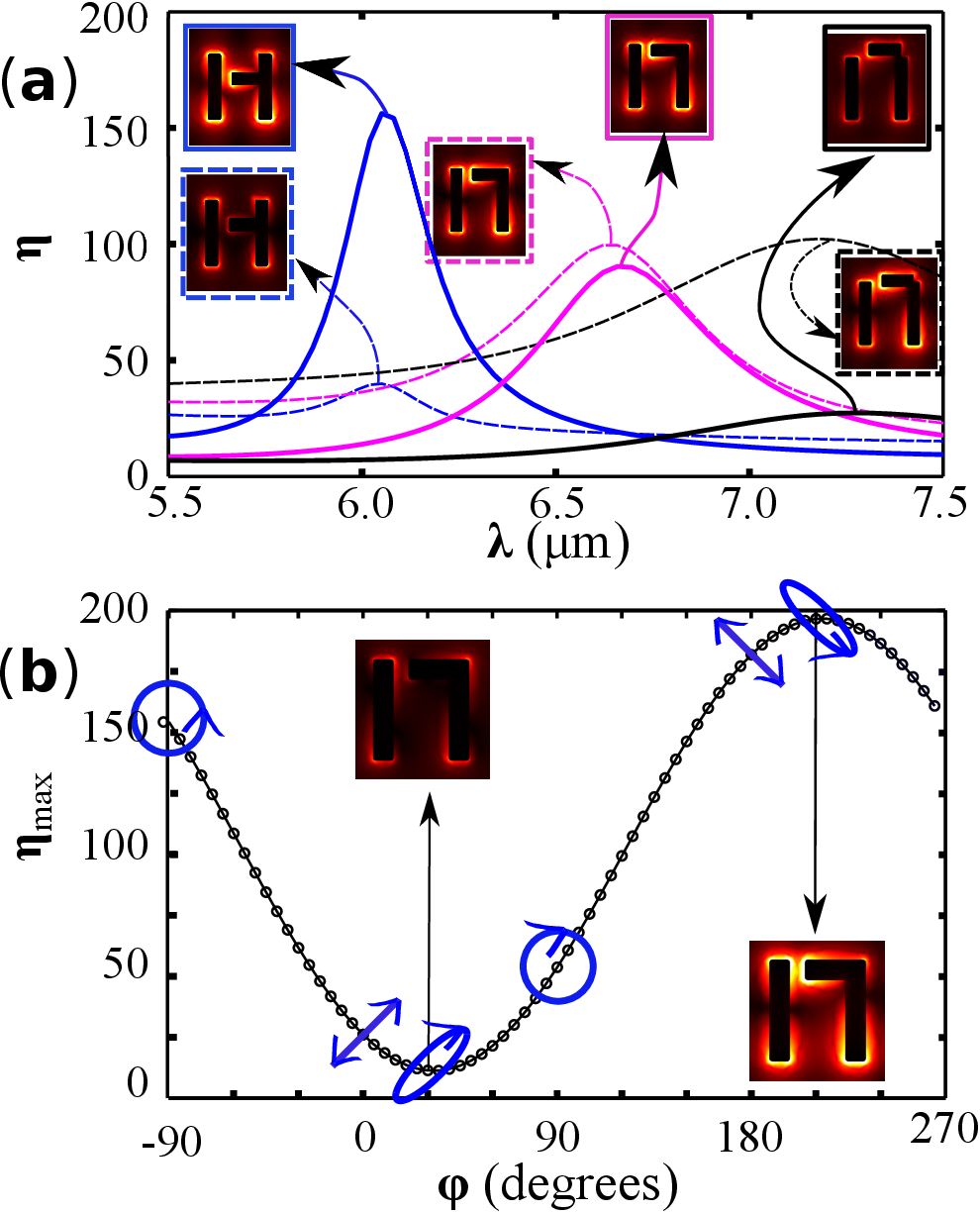} \\
\caption{(color online) COMSOL simulations of the optical intensity enhancement $\eta = \langle |\vec{E}|^2 \rangle/E_{0}^2$ averaged over all (except metal-substrate) metal interfaces in plasmonic DCF metasurfaces. (a) $\eta$ for $x-$polarized (solid) and $y-$polarized (dashed lines) incident light for weak-coupled ($s_y=0.32\mu$m), intermediately-coupled ($s_y=0.72\mu$m), and strongly-coupled ($s_y=1.14\mu$m) regimes. Insets: near field profiles. (b) The dependence of $\eta_{\rm max}$ on the incident light chirality parametrized by the phase shift $\phi$ (see text) for the $s_y=0.52\mu$m metasurface. Insets: near-field intensities for $\phi=30^{\circ}$ (left) and $\phi=210^{\circ}$ (right). Closed directional loops indicate the polarization state of incident light as a function for $-90^{\circ} < \phi < 270^{\circ}$.}\label{fig:quad}
\end{figure}

{\it Circularly dichroic energy concentration}-- It has been long
recognized that one of the attractions of Fano-resonant
metasurfaces is strong optical field enhancement due to the
excitation of dark plasmonic resonances. Controlled by the
metasurfaces' geometry, those can be utilized for
sensing and nonlinear applications. As we demonstrate below, optical field concentration can also be
controlled through interference~\cite{stone_prl10} by engineering a general elliptic polarization state of the incident light. The consequences of this effect include giant CD in absorption and transmission~\cite{zheludev_prl06}. 

First, we calculate the amplitude $Q$ of the dark resonance in the proximity of the DQ mode in response to the incident light with complex-valued amplitudes $(E_x,E_y) = E_0(1,\exp{i\phi})/\sqrt{2}$. Here $\phi$ is the relative phase between the $x-$ and $y-$ polarizations that determines the state of elliptic polarization as illustrated in Fig.~\ref{fig:quad}(b). At the Fano resonance, the amplitude of the DQ mode is expressed from Eq.~(\ref{eqn:cmt2}) as $Q=E_0 \left( q_{xx} + \exp{(i\phi)} q_{yy} \right)/(\omega-\tilde{\omega}_Q^{\prime})$, where
\begin{equation}\label{eq:field_enhancement}
    q_{xx} = \frac{\alpha_x \ \kappa_{XQ}(s_y)}{\tilde{\omega}_X - \tilde{\omega}_Q},
    q_{yy} = \frac{\alpha_y \ \kappa_{XQ}(s_y) \ \kappa_{XY}(s_y)}{(\tilde{\omega}_X - \tilde{\omega}_Q)(\tilde{\omega}_Y - \tilde{\omega}_Q)}
\end{equation}
are complex-valued field enhancement coefficients.

Because of the different scaling of $q_{xx}$ and $q_{yy}$ with
$s_y$, it follows from Eq.~(\ref{eq:field_enhancement}) that in
the weak (strong) coupling regime the strongest field enhancement is achieved for the $x-$ ($y-$) polarized light. This is confirmed by electromagnetic COMSOL simulations of the surface-averaged optical intensity enhancement $\eta \equiv \langle |\vec{E}|^2 \rangle/E_0^2$ shown in Fig.~\ref{fig:quad}(a) for the $x-$ and $y-$polarized light incident on DCF metasurfaces with different values of $s_y$. An even more remarkable conclusion derived from Eq.~(\ref{eq:field_enhancement}) is the possibility of controlling the field enhancement using elliptically-polarized light.

The resulting CD (i.e. the dependence of $\eta
\propto |Q|^2$ on $\phi$ ) is the consequence of the finite phase
difference $\Delta \phi = \arg{\left(q_{xx}/q_{yy}\right)}$: the field enhancement is maximized (minimized) for $\phi_{\rm max} = \Delta \phi$ ($\phi_{\rm min} = \Delta \phi + \pi$) due to the
constructive (destructive) interference between the two excitation pathways. Moreover, symmetry properties of the coupling
coefficients  $\kappa_{XQ}$ and $\kappa_{XY}$ ensure the following property of the structure's enantiomeric partner produced by displacing of the HCN to the opposite side ($s_y \rightarrow -s_y$) of the symmetry plane: $|Q|(\phi,-s_y)=|Q|(\phi+\pi;s_y)$.

These qualitative results are confirmed by COMSOL simulations carried out for the $s_y=0.52\mu$m structure chosen because $|q_{xx}| \approx |q_{yy}|$ according to Fig.~\ref{fig:quad}(b). The peak enhancement factor $\eta_{\rm max}(\phi) \equiv \eta(\lambda_{DQ}^{\prime},\phi)$ plotted in Fig.~\ref{fig:quad}(b) reveals strong CD. The ratio of the highest ($\phi_{\rm max}\approx 210^{\circ}$) to lowest ($\phi_{\rm min}\approx 30^{\circ}$) near-field intensities is $\approx 20$, and the enhancement ratio for right-hand circularly polarized (RCP) light ($\phi=-90^{\circ}$) is three times higher than for the left-hand circularly polarized (LCP) light ($\phi=90^{\circ}$).

Because $\eta_{\rm max}(\phi)$ and peak Ohmic loss $A(\phi)$ are proportional to each other, $\phi$ emerges as a  powerful tool for controlling the absorption using the interference of the polarization components of elliptically polarized light. For example, we find for the DCF metasurface shown in Fig.~\ref{fig:quad}(b) that $A(\phi_{\rm min}) \approx 1.5\%$, i.e. the absorption is coherently suppressed. On the other hand, $A(\phi_{\rm max}) \approx 41\%$, i.e. the absorption is coherently enhanced almost to the absolute absorption limit ($A_{lim}=50\%$)of a thin layer in vacuum~\cite{abajo_rmp07}. Another important consequence of the related giant CD absorbance of the DCF metasurface ($A_{\rm RCP} = 30\%$ vs $A_{\rm LCP} = 13\%$) is that it exhibits circular conversion dichroism~\cite{zheludev_prl06} due to the dissipation asymmetry between RCP and LCP light.

In conclusion, we have demonstrated that a quantum mechanical
phenomenon of double continuum Fano (DCF) interference can be
classically emulated using an asymmetric plasmonic metasurface.
The relative coupling strength between the discrete and two
continuum states, distinguished from each other by their
polarization, were experimentally varied by changing the degree of symmetry breaking of the metasurface. The phenomenon of continuum
state competition, by which one of the continuum states suppresses the Fano resonance for the orthogonal light polarization, was
observed and analytically explained. This work opens new
possibilities for controlling optical energy concentration on the
nanoscale using the chirality of the incident light, thereby
providing a simple and powerful tool for developing novel nanophotonic applications such as sensors and detectors.

This work was supported by the Office of Naval Research (ONR) Award N00014-13-1-0837, by the National Science Foundation(NSF) Award DMR 1120923, and by US Army Research Office (ARO) Award W911NF-11-1-0447.

\end{document}